\documentstyle[preprint,aps]{revtex}

\begin{document}

\draft

\title{Cancer Detection via Determination of Fractal Cell Dimension}

\author{Wolfgang Bauer}
\address{Department of Physics and Astronomy and
National Superconducting Cyclotron Laboratory, Michigan State
University, East Lansing, MI 48824-1321, USA}

\author{Charles D. Mackenzie}
\address{Department of Pathology, Michigan State University,
East Lansing, MI 48824, USA}

\maketitle

\begin{abstract}
We utilize the fractal dimension of the perimeter surface of cell sections as
a new observable to characterize cells of different types.
We propose that it is possible to distinguish cancerous from healthy
cells with the aid of this new approach.  As a first application we
show that it is possible to perform this distinction between patients
with hairy-cell lymphocytic leukemia and those with normal blood
lymphocytes.
\end{abstract}


In the final diagnosis of neoplasia the pathologist generally relies on
the qualitative and empirical parameters of the cells in biopsies
or cytological preparations.  These experiential approaches have been
aided by morphometric methods, such as determination of surface area,
volume, axes ratios, estimation of population density, and other
methods derived from basically Euclidian geometry \cite{AD82}, to enhance the
examination of the internal components such as the nucleus, number of nucleoli,
amount of chromatin, nuclear membrane abnormalities, together with
perturbation of the cytoplasmic membrane, and degree of differentiation.

In mathematics and the physical sciences, it has been known since the turn
of the century that classical Euclidian concepts, such as perimeter length,
surface area, or volume, do not yield finite answers for certain objects.
The earliest and probably most famous example is Koch's Snowflake
\cite{Koc04} with infinite perimeter surface length and finite area.
A similar problem is encountered when trying to measure the coastal length of a
country like Great Britain, for example.  There one finds that the answer
depends on the size $\epsilon$ of the yardstick one is using, and that it
diverges to infinity as the size of the yardstick approaches 0 \cite{Ric61}.
In order to describe this kind of complexity found in nature, Mandelbrot
developed the concept of fractal geometry \cite{Man77}.  He introduced the
fractal dimension as a more convenient ways to parametrize the surface,
$L(\epsilon) \sim F\, \epsilon^{1-d}$, where $d$ is the fractal Haussdorff
dimension.

In biology and medicine, fractals are just now beginning to have first
applications \cite{Wes90,Bun94}.
Recently studied have been the shape of neurons in
vertebrate central nervous systems \cite{Smi89},
the frequency spectrum in human heartbeats
\cite{Pen93},
vascular and airway systems \cite{Gan1993}, brain surface \cite{Priebe1994},
trabecular bone \cite{Buckland1994} and pathological entities
such as colorectal polyps \cite{Cross1994}, or
epithelial tissues in the oral cavity \cite{Landini1993}.

Fractals have also been utilized at the cell and sub-cellular level.
The anatomical outline of neuronal cells \cite{Traverso1992},
such as glia \cite{Smith1994} and retinal cells \cite{Holb1994}
have been compared, and so has the nuclear
outline of cervical epithelial cells \cite{MacAulay1990} for cancerous
changes.  Surface receptor localization \cite{Sadana1994}
and changes in cytoplasmic components such as cytosketon \cite{Aon1994}
and the nucleus of other cell types has revealed functional
correlates.

In this current study we focus on the use of fractals at the
cellular level.
In the following we describe our method of extraction of fractal
dimensions for biological cell samples \cite{BM94}.
We illustrate the method by
showing the results of the different steps on a human lymphocyte
affected by hairy-cell leukemia.  However, the method is general and
can be used on many types of cells or organisms such as
protozoa and bacteria.

We start by obtaining an electron microscope image of
a cell section (Fig.\ 1a).
This image is digitized in 256 gray-levels at fairly high
pixel resolutions.  The image shown in Fig.\ 1a contains $750\times 675$
pixels.
By compiling grey level distribution histograms, one is able to automatically
set grey level thresholds for the cell membrane and the cell nucleus.  In
Fig.\ 2 we show the grey level distribution for the digitized image of
Fig.\ 1a.
There are two local minima visible.  The lower of these two minima corresponds
to the threshold for the cell membrane, the object of interest here.
The long vertical line corresponds to a grey level of 80, chosen for this
particular cell.  However, the exact choice (within $\pm$ 5) of the threshold
value is relatively unimportant.  We have investigated the sensitivity of our
final results to the threshold value and find only a weak dependence over this
range.

In the next processing step, every pixel with grey level above the threshold
value is assigned black, and everyone below white.  This results in the
picture shown in Fig.\ 1b.  One can now already see a rough image of the cell
section's surface.  However, there are small inclusions inside the cell which
now appear as white spots, but upon examination of Fig.\ 1a clearly should
belong to the cell.  In the same way, there
is extraneous extracellular material (seen as black
aggregates),
mainly from proteins in the solution, which clearly don't belong to
the cell, either.  These small black and white spots have to be removed for
us to obtain meaningful information on the surface of the cell.

Conventional speckle removal algorithms do not work in our case, because
techniques like neighboring-pixel-averaging tend to smoothen the true cell
surface as well, often resulting in a changed fractal dimension.
Fortunately, we were able to utilize a cluster-recognition
algorithm, developed by one of us to identify fragments in nuclear \cite{Bau85}
and bucky-ball \cite{Bau94} fragmentation.
This is yet another example where basic research in one field has lead
to a completely unexpected applied spinoff in another.
With this algorithm we identify all clusters below a minimum size and invert
their grey level value.  This minimum size was
universally chosen to be 300 pixels in all cells processed by us.
We have varied this parameter and find only minimal influence
on our final results over a reasonable range (100 -- 500).
The output of this algorithm is shown in Fig.\ 1c.  One can clearly see that
all speckles have been successfully removed, and that the overall shape of the
surface was not altered in the process.

It is now straightforward to obtain pixels on the surface of the cell section
by taking partial derivatives in the $x$ and $y$ coordinates.  These
derivatives are surface-peaked and have a value of 0 elsewhere.  The result
of this procedure is shown in Fig.\ 1d, where the surface is now represented
as a black line on an otherwise white background.  By comparing Fig.\ 1d with
Fig.\ 1a, one can see that the automated surface recognition procedure outlined
here does the job rather well.

We now determine the fractal dimension, $d$,
of the surface by utilizing the box-counting method.  It was shown
that this method yields quantitative agreement with the line-segment
method discussed above \cite{Fed88}.
In the box-counting method, one utilizes the definition
of the Hausdorff-dimension, where the number, $N(\xi)$, of squares of
side-length $\xi$ needed to cover a set increases like
$N(\xi) \propto \xi^{-d}$ for $\xi\rightarrow 0$
for a set of fractal Hausdorff dimension $d$.  One then can theoretically
obtain the fractal dimension from
$d = - \log[N(\xi_1)/N(\xi_2)]/\log[\xi_1/\xi_2]$.

However, in practice there are some difficulties associated with arbitrarily
choosing $\xi_1$ and $\xi_2$.  We chose our sidelength as integer powers
of 2, $\xi_l = 2^{-l}$.  For a too small value of $l$, $N(\xi_l)$ will
increase like for a 2-dimensional set.  For a too large value of $l$,
$N(\xi_l)$ will approach a constant, the total number of pixels on the
surface.  This asymptotic limit for sets of finite numbers of points has
been studied, and methods have been devised to correct for this effect
\cite{Gra83}.  However, in our case an approximate determination of
$d$ within $\pm$ 0.03 is fully sufficient, and
we found that a linear regression fit for the $l$-values in the
range 3 to 6 (corresponding to a $\xi$-range between 1/8 and 1/64 and
total numbers of boxes between 64 and 4096) yielded
satisfactory results.  In all cases considered, we obtained correlation
coefficients of $1-r<10^{-2}$.  We also tested our method on known fractals
such as Koch's Snowflake and Sierpinsky's Carpet, and obtained the exact
results to within the uncertainty specified above.

In Fig.\ 3, we show (circles) the results of our box-counting algorithm for the
surface of the cell section as displayed in Fig.\ 1d.  The total number of
surface pixels happened to be 5297 in this case, a limit reached for $l = 10$.
The power-law fit in the $l$-interval from 3 to 6
yields a fractal dimension of $d=1.34$ (with a
correlation coefficient of $r=0.99975$) and is represented by the straight
line.

With this method of determination of the fractal box-counting dimension
of the surface of a cell section it is possible to derive a {\em quantitative}
measure for the raggedness of cells or small biological organisms.  We
report here on our initially most important result, the investigation
of human blood lymphocytes affected by hairy-cell leukemia.

To perform our test, we obtained archive electronmicrographs of white blood
cell preparations of two patients who died
of hairy-cell leukemia.  As controls,
we also processed white blood cells of healthy patients.  In each group
patient we examined samples of about 100 cells, for which we determined the
fractal dimension individually.
The result of applying our method to these cell samples is shown in
Fig.\ 4.  The number of cells with fractal dimension $d_{\rm eff}$ are
binned in (area normalized)
histograms and displayed as a function of $d_{\rm eff}$.

In Fig.\ 4a we show the distribution of white blood cells for the healthy
patients.
Hairy-cell leukemia is a form of lymphocytic neoplasia that produces varying
proportions of circulating neoplastic cells characterized by their
morphologically altered cytoplasmic membranes, which carry considerably
more surface projections (pseudopods) than do normal healthy lymphocytes.
However, other normal cells in blood, e.g.\ neutrophils and other granulocytes,
which may constitute 50\% or more of the sample, can also have a rough
(``hairy'') surface.  One can discriminate against
them by examining deformation of the cell nucleus.  Taking this criterion into
account, one arrives at the histogram shown in Fig.\ 4b, the distribution
of fractal dimensions for healthy lymphocytes.

In Fig.\ 4c we show the distribution obtained from the lymphocytes
of the cancer patients.  The dotted vertical
line indicates a fractal dimension of 1.28.  None of the cells of the healthy
patients had a fractal dimension larger than this value, whereas a large
percentage of the cells of the cancer patients had $d>1.28$.
The difference between the fractal dimension distributions for
healthy and hairy-cell leukemic patient lymphocytes is obvious.

In summary,
we have used the concept of fractal dimensional analysis of surfaces
of individual cells.  We have shown
on the example of a white human blood cell how our method works, and that a
meaningful extraction of a fractal dimension is possible by using automated
computer procedures.  In this
first exploratory study we have presented a classification of cell samples
in terms of a fractal dimension histogram of cells.  We showed that it is
possible to distinguish between healthy persons and hairy-cell leukemia cancer
patients on the basis of our method.

In the future we see two direct application of our method.  First, one can
envision our system as a diagnostic aid for the practicing pathologist.  With
it a technician can use our system on a very large cell sample,
picking out only the most interesting diagnostically relevant individual
cells for visual inspection by the pathologist.  This is possible at the
present stage of development.
A second, completely automated, application should also be possible.
For this, however, carefully controlled, double-blind
clinical studies are necessary.  We plan to start this in the near future.

We expect that our method will also be readily applicable to
a host of other cancers,
most prominently breast cancer.  We speculate that there might exist a
connection between shape change, i.e.\ change in fractal dimension, and
metastasis.  If this holds true, then our method could be used for early
detection.

\clearpage

\subsection*{Figure Captions}

\begin{description}

\item[Fig.\ 1] a) Electron microscope image of a section of a human lymphocyte,
      affected with hairy-cell leukemia, digitized with 256 grey levels.
      b) Black/white representation of (a) with a grey level
      threshold set at 80.
      c) Image after application of our small-cluster
      removal algorithm.
      d) Perimeter surface of the image, obtained by taking
      derivatives in x and y directions.
\item[Fig.\ 2] Grey level distribution for image shown in Fig.\ 1a.
\item[Fig.\ 3] Number of boxes touched by the perimeter surface of Fig.\ 1d.
      as a function of $l$, where the box side length is given by $s=2^{-l}$.
      The straight line is a fit in the interval $l=[3,6]$.
\item[Fig.\ 4] Histograms of fractal dimensions of sections of individual
      cells:  a) White blood cells of healthy persons; (b) Lymphocytes of
      healthy persons; (c) Lymphocytes of hairy cell leukemia patients.
      The dotted vertical line indicates a fractal dimension
      of 1.28, a value which is surpassed only by cells from the cancer
      patients.

\end{description}

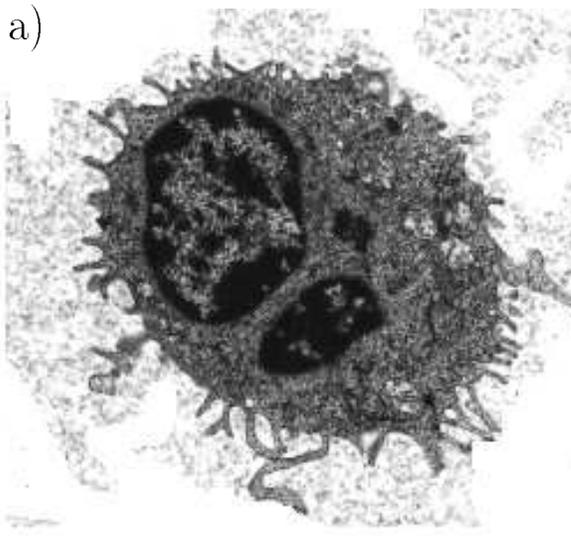
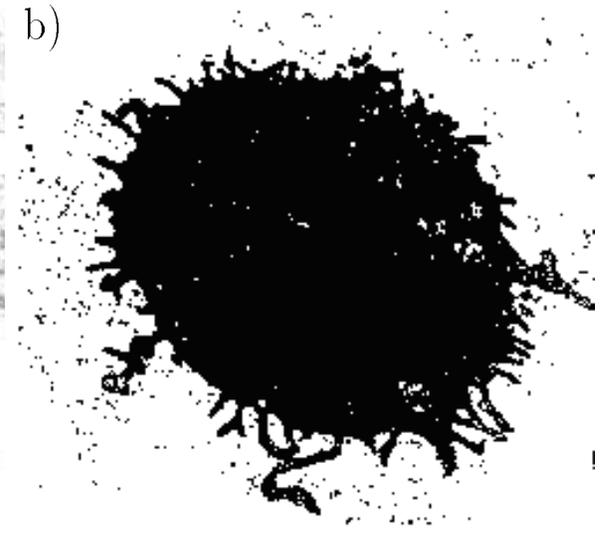
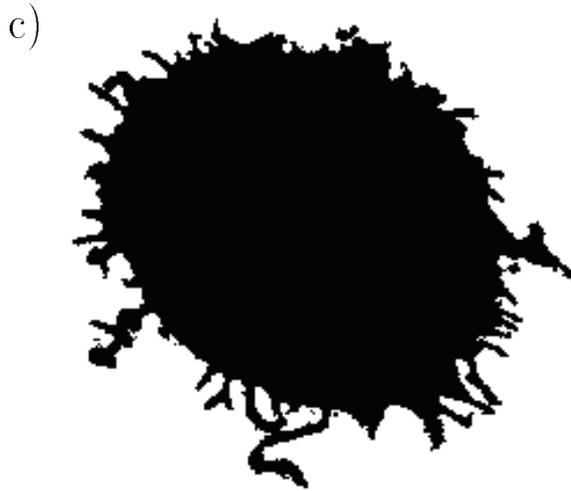
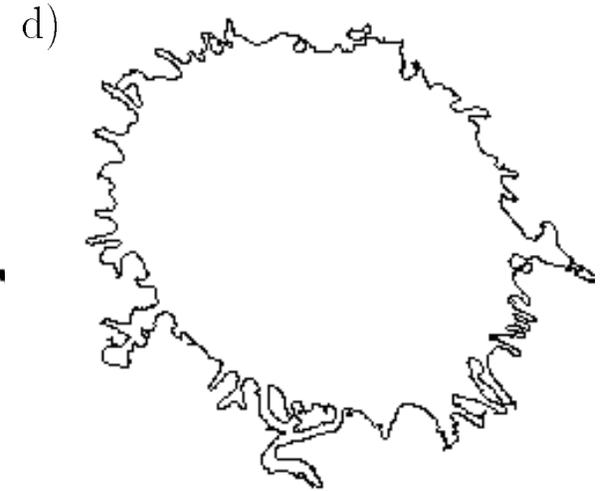
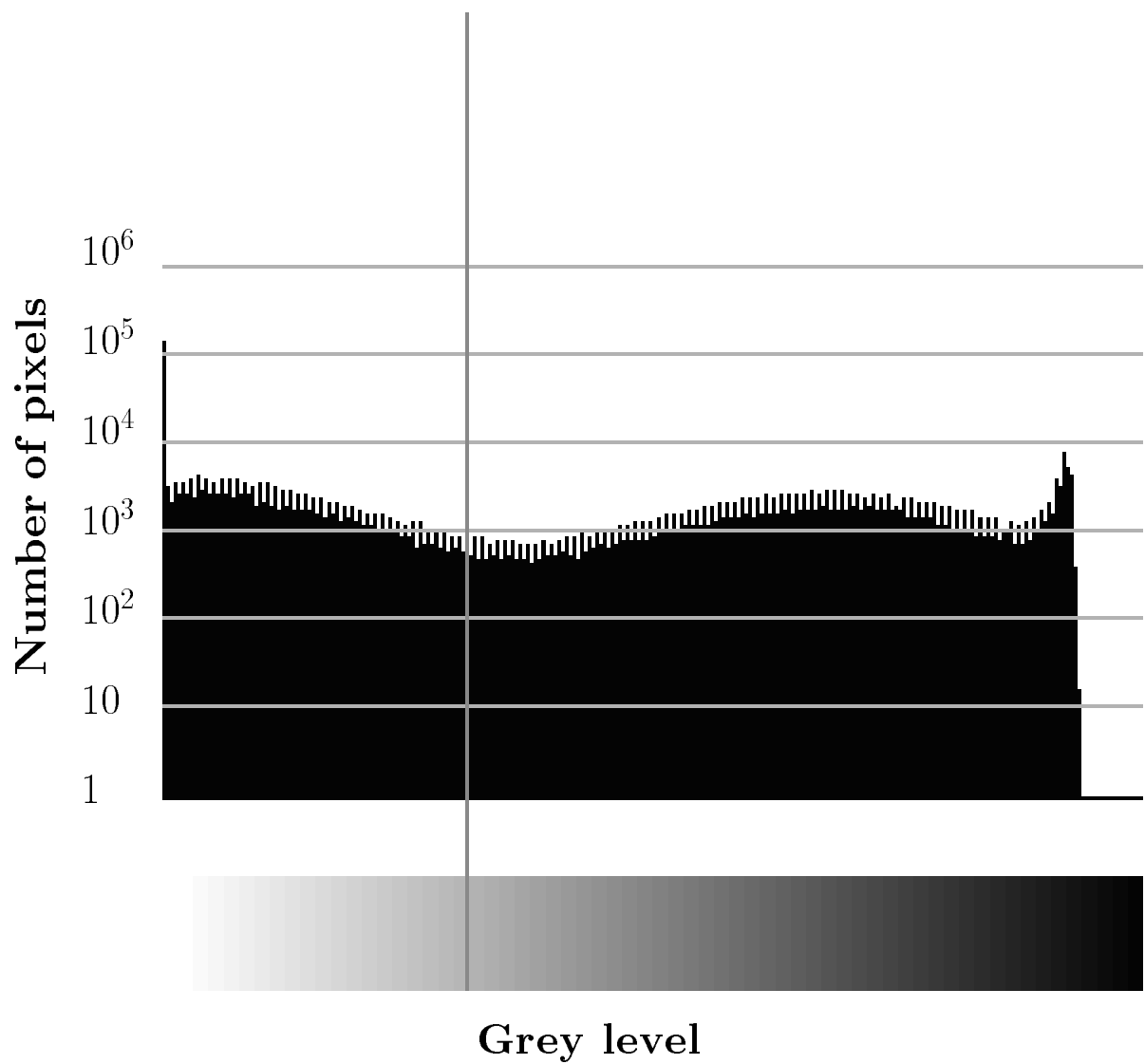
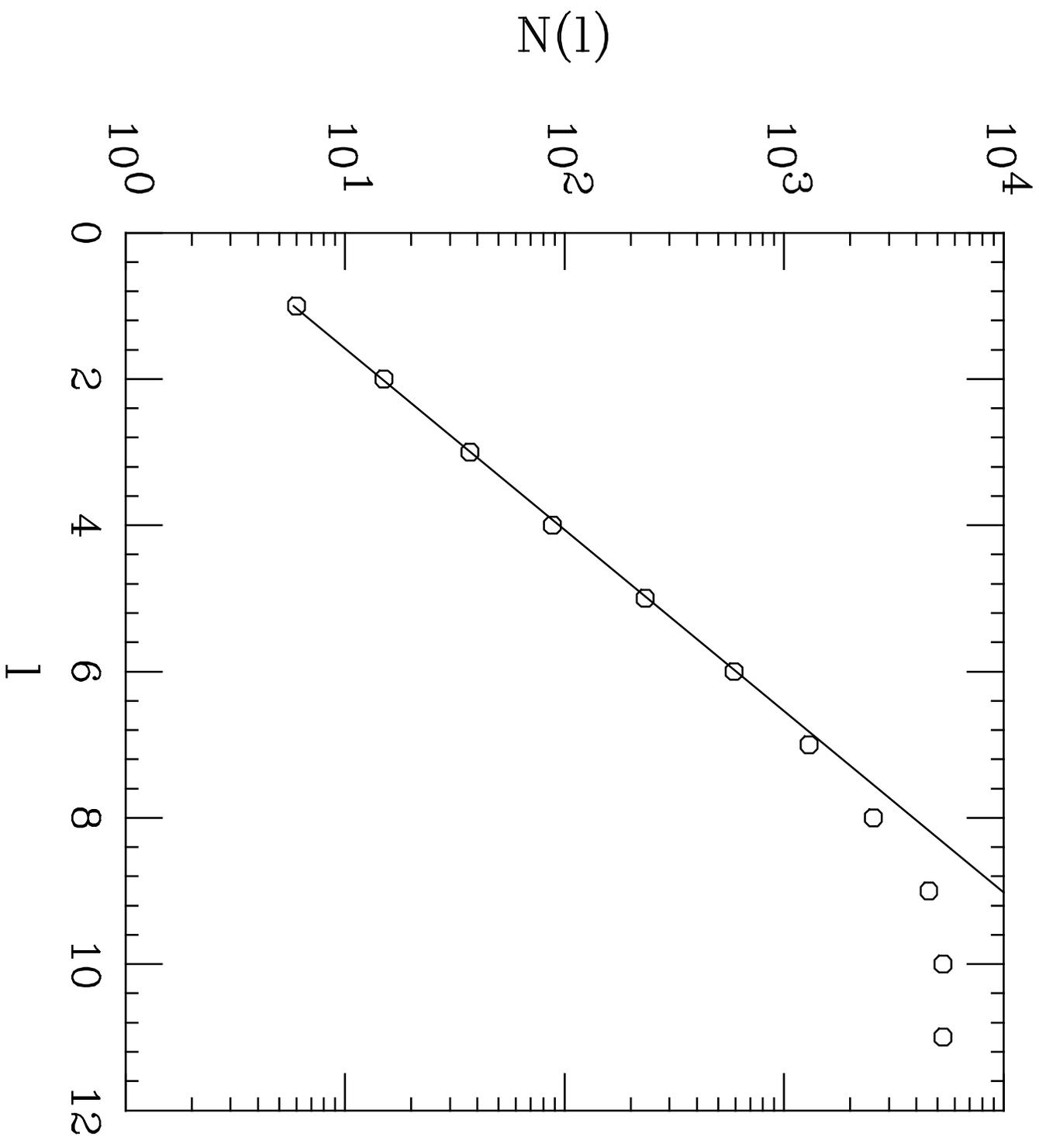
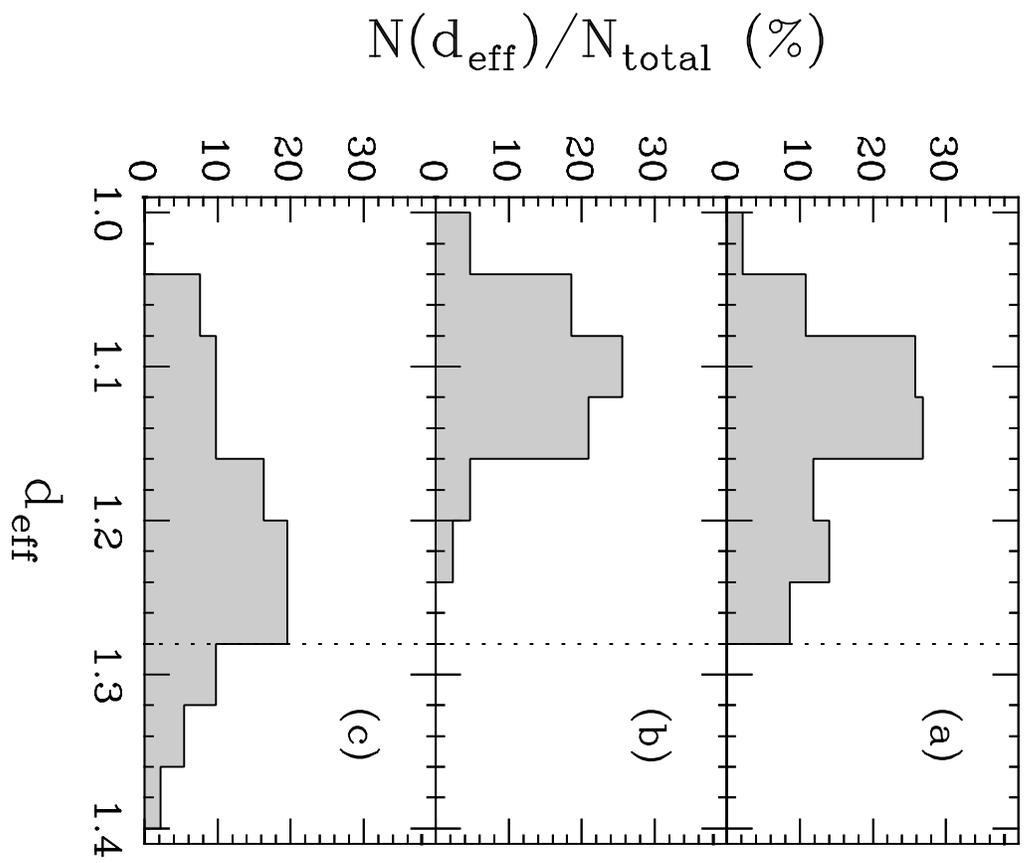
\end{document}